\documentclass[
reprint,
superscriptaddress,
 amsmath,amssymb,
aps,
prb, 
]{revtex4-2}
\DeclareMathOperator{\Tr}{Tr}
\usepackage{graphicx}
\usepackage{float}
\usepackage{dcolumn}
\usepackage{bm}
\usepackage[colorlinks = true,
linkcolor = blue,
urlcolor  = blue,
citecolor = blue,
anchorcolor = blue]{hyperref}
\usepackage{comment}
\usepackage{xcolor}

\begin{document}

\preprint{APS/123-QED}

\title{Dynamics of defects and interfaces for interacting quantum hard disks}

\author{Fabian Ballar Trigueros}
\altaffiliation{These authors contributed equally to this work.}
\affiliation{%
 Theoretical Physics III, Center for Electronic Correlations and Magnetism,
Institute of Physics, University of Augsburg, 86135 Augsburg, Germany}%

\author{Vighnesh Dattatraya Naik}
\altaffiliation{These authors contributed equally to this work.}
\affiliation{%
 Theoretical Physics III, Center for Electronic Correlations and Magnetism,
Institute of Physics, University of Augsburg, 86135 Augsburg, Germany}%

\author{Markus Heyl}
\affiliation{%
 Theoretical Physics III, Center for Electronic Correlations and Magnetism,
Institute of Physics, University of Augsburg, 86135 Augsburg, Germany}%
\affiliation{Centre for Advanced Analytics and Predictive Sciences (CAAPS), University of Augsburg, Universitätsstr. 12a, 86159 Augsburg, Germany}

\date{\today}

\begin{abstract}
\noindent 
Defects and interfaces are essential to understand the properties of matter. However, studying their dynamics in the quantum regime remains a challenge in particular concerning the regime of two spatial dimensions. Recently, it has been shown that a quantum counterpart of the hard-disk problem on a lattice yields defects and interfaces, which are stable just due to quantum effects while they delocalize and dissolve in an analogous classical stochastic process. Here, we study in more detail the properties of defects and interfaces in this quantum hard-disk problem with a particular emphasis on the stability of these quantum effects upon including perturbations. Specifically, we introduce short-range soft-core interactions between the hard disks. From both analytical arguments and numerical simulations we find that large classes of defects and interfaces remain stable even under such perturbations suggesting that the quantum nature of the dynamics exhibits a large range of robustness. Our findings demonstrate the stability and non-classical behavior of quantum interface dynamics, offering insights into the dynamics of two-dimensional quantum matter and establishing the quantum hard-disk model as a platform for studying unconventional constrained quantum dynamics.
\end{abstract}

\maketitle


\section{Introduction}

Interfaces in two-dimensional quantum systems can exhibit remarkably rich dynamics driven by quantum interference and constraints. 
Among the most striking phenomena are interface-driven phase transitions that are decoupled from the bulk behavior, such as the roughening transition, where interfaces change from smooth to rugged configurations~\cite{Beijeren1975InterfaceSI,PhysRevB.28.5338}. 
These interfaces also display complex real-time dynamics, including non-trivial relaxation and signatures of non-ergodicity such as quantum many-body scars and Hilbert space fragmentation~\cite{Balducci_2023, PhysRevLett.129.120601,Krinitsin2024}. 

Recent works have further expanded this landscape by identifying a mechanism of ergodicity breaking in constrained quantum systems, namely quantum many-body caging ~\cite{benami_2025,tan2025,jonay2025}. Here, quantum interference and local constraints cooperate to dynamically confine the system to a subset of configurations that form flat bands in the spectrum. 
These bands are associated with compact localized many-body eigenstates, which cannot hybridize with the rest of the Hilbert space due to destructive interference. 
As a result, certain initial states exhibit persistent memory and fail to thermalize, even in the absence of disorder or fine-tuning.

Despite these theoretical advances, simulating real-time evolution in 2D quantum many-body systems remains a formidable challenge. 
However, constrained quantum systems provide a promising route forward: the presence of local constraints significantly reduces the size of the dynamically accessible Hilbert space, making them amenable to both theoretical analysis and experimental simulation. 
In particular, Rydberg atom arrays, which naturally implement kinetic constraints through the Rydberg blockade mechanism, have emerged as an ideal platform for studying such dynamics~\cite{Weckesser2024bpd,Bernien_2017,Keesling2019,deLeseleuc2019,Browaeys_2020,Ebadi2021,Semeghini2021,Scholl_2021,Bluvstein2021}.

In our previous work~\cite{QHD}, we introduced and studied the quantum hard-disk model, a lattice-based quantum analog of the classical hard-disk system. 
Here, the exclusion constraint is realized via particles that carry a finite “excluded volume,” implemented as a hard-core radius around each particle (see Fig.~\ref{fig:figure_0}(A)). While the model’s equilibrium properties matched the classical case—featuring a crystalline phase at high density—its dynamics revealed intrinsically quantum phenomena. Most notably, we identified a large number of defect and interface configurations that are dynamically stable only in the quantum model, owing to interference effects that prevent thermalization. 
As a result, the model exhibits nonergodic behavior even in two dimensions, defying conventional expectations for generic interacting systems. Recent developments have linked these nonergodic structures to the mechanism of quantum many-body caging, where constraints and interference stabilize long-lived atypical eigenstates~\cite{benami_2025}.

While our prior work established the existence of stable quantum interfaces, a key open question remained: How robust are these structures under perturbations or deformations of the Hamiltonian? 
To answer this, we now extend our study by introducing additional short-range soft-core interactions, probing whether nonergodic behavior and defect stability persist beyond this limit. So this model is a compelling platform for studying nonergodicity, and they exhibit rich behavior. Moreover, a one-dimensional analog of the model, involving quantum hard rods, has now been realized experimentally~\cite{Weckesser2024bpd}, further motivating the study of this model.

We find that these interactions induce a form of dynamical heterogeneity, where different initial configurations exhibit markedly different relaxation dynamics. 
Some interface-like structures remain nonthermal for arbitrarily long times, retaining memory of their initial state. 
This robustness is tied to the persistence of quantum many-body cages, which survive even in the presence of interactions. 
Other configurations, by contrast, show long-lived plateaus followed by eventual thermalization, while some states exhibit coexistence of ergodic and nonergodic regions, revealing spatially resolved relaxation dynamics.

These findings highlight the quantum hard-disk model as a robust platform for exploring constrained quantum dynamics and nonergodicity in 2D, with relevance for current quantum simulation experiments and broader questions about the emergence of thermalization.

This paper is organized as follows: In Section~II, we introduce the model and the specific deformations we consider. 
Section~III summarizes our previous results on constrained dynamics and Hilbert space fragmentation. 
Section~IV presents a detailed analysis of the dynamical behavior of point-like and interface-like defects. 
Finally, Section~V discusses implications and possible directions for future research.

\begin{figure*}
    \centering
    \includegraphics[width=\linewidth]{figure_-2.pdf}
    \caption{Defect dynamics on the square lattice with interactions. (\textbf{A}) Schematic representation of the interacting quantum hard disk model on a square lattice. The interactions are of next-to-nearest-neighbor type, acting only between particles located diagonally. There are no interactions between particles that are separated by two sites in the same row or column. (\textbf{B}) The Hilbert space fragmentation as a function of particle density $\eta$. Here, $\eta_1 = 1/L$, $\eta_2 = \frac{1}{2} - \frac{\lceil L/2 \rceil }{ L^2}$, and $\eta_{\textrm{max}} = \frac{1}{2} - \frac{L(\textrm{mod} 2)}{2L^2}$. $\eta^*$ denotes the threshold density for the weak-to-strong crossover region. (\textbf{C}) Long-time dynamics of on-site occupations at $Jt = 10^{2}$, showing memory of the initial state even in the presence of interactions.}
    \label{fig:figure_0}
\end{figure*}

\section{Model} 
To model interacting quantum hard disks on a square lattice, we consider a system of hard-core bosons described by a Hamiltonian that includes nearest-neighbor hopping and next-to-nearest-neighbor interactions:  
\noindent
\begin{equation}
    H = J\sum_{\langle i, j \rangle}  P_i {a}_i^{\dagger}  a_j P_j + P_j a_j^{\dagger}  a_i P_i + \lambda \sum_{\langle\langle i, j \rangle\rangle} n_i n_j \, \, .
    \label{eq:Hamiltonian}
\end{equation}
\noindent
Here, $a_i^{\dagger}$ and $a_i$ are the creation and annihilation operators for a hard-core boson on lattice site $i = 1, \dots, L^2$, where $L^2$ is the total number of lattice sites, and the lattice constant is $a=1$. 
The lattice has open boundary conditions, ensuring that hopping and interaction terms are excluded for sites beyond the boundaries.  

The excluded volume effect due to the hard disks is implemented using projection operators  $ P_i =  \prod_{j \ni |\vec r_i - \vec r_j| = 1} (1 - n_j),$ which prevent particles from occupying nearest-neighboring sites (see Fig.\,\ref{fig:figure_0}\textbf{A} for an illustration).
Here, $n_i = a_i^{\dagger} a_i$ represents the occupation number operator. 
These constraints significantly reduce the computational basis size by excluding states that violate the hard-disk condition, making numerical simulations more efficient, see for instance Fig.~\ref{fig:figure_0} for a $20\times20$ lattice.
The terms $\langle i,j \rangle$ and $\langle\langle i,j \rangle\rangle$ represent nearest-neighbor and next-to-nearest-neighbor pairs, respectively. Here, $\lambda$ denotes the strength of the next-to-nearest-neighbor interactions, while $J$ specifies the hopping strength, which is set to $J = 1$ in our analysis.
The model in Eq.~(\ref{eq:Hamiltonian}) exhibits a $U(1)$-symmetry in terms of particle number conservation.
In the following, we will therefore typically fix the particle density $\eta = M / L^2$ with $M$ denoting the total number of particles.
This model is, in principle, experimentally realizable, as demonstrated for the $1$D case in \cite{Weckesser2024bpd}.
Of particular importance in this context are Rydberg atomic systems: the Rydberg blockade mechanism automatically enforces the excluded volume constraint by preventing the occupation of neighboring sites. 
The case of free hard disks ($\lambda = 0$) was analyzed in detail in our previous work \cite{QHD}.
In this study, we extend our investigation by examining the robustness of these results in the presence of the additional soft-core interactions between the hard disks at $\lambda\not=0$.
 
\section{Hilbert-Space fragmentation}
A key characteristic of strong local constraints is the fragmentation of the Hilbert space into kinetically disconnected sectors.
As demonstrated in our previous work, Hilbert-space fragmentation~\cite{Pai2019,Sala2020,Rakovszky2020,Khemani2020,Moudgalya2022,Smith2017,Smith2017_2,Brenes2018,Russomanno2020,Karpov2021,wang2024,Chakraborty2022,Chakraborty2022_2} plays a fundamental role in governing the dynamics of defects and interfaces in systems of free hard disks~\cite{QHD}.
A summary of these findings is provided in the following section. 
Notably, the introduction of soft-core interactions does not alter the underlying fragmentation structure of the Hilbert space.

Hilbert-space fragmentation can be classified into two distinct regimes: strong and weak fragmentation.  
The fragmentation regime is determined by how the ratio between the size of the largest fragment, $\mathcal{N}_{\textrm{max}}$, and the total Hilbert-space dimension, $\mathcal{N}$, scales in the thermodynamic limit. 
Strong fragmentation occurs when $\mathcal{N}_{\textrm{max}}/\mathcal{N} \to 0$, indicating a highly disconnected Hilbert space.  
Weak fragmentation is characterized by $\mathcal{N}_{\textrm{max}}/\mathcal{N} \to 1$, where most states remain kinetically accessible, while some disconnected states exist~\cite{Moudgalya2022, kwan2023minimal}.

The studied system exhibits both weak and strong fragmentation depending on the particle density $\eta$. In Fig.\,\ref{fig:figure_0}\textbf{(B)}, we present a refined analysis of the fragmentation properties identified in our previous work as a function of $\eta$.

The key structural element in our basis configurations that governs fragmentation in our model is what we term a snake—a fully filled main diagonal or its anti-diagonal within the configuration.
The presence of a snake acts as a barrier, effectively dividing the lattice into two distinct regions and preventing particle movement between them, thus leading to fragmentation.
At low particle densities, no snakes are present, and therefore, fragmentation does not occur.
As the particle density $\eta$ increases to $1 / L$, configurations can contain snakes.
Beyond this threshold, the Hilbert space fragments into several smaller components, while a dominant large fragment persists, leading to weak fragmentation.
As $\eta$ increases to $\eta > (1/2) - \lceil L/2 \rceil / L^2$ (where $\lceil L/2 \rceil$ denotes the ceiling function), each allowed configuration must include at least a snake or a fully frozen parallel diagonal. 
These frozen diagonals effectively separate the system into two distinct regions.
Consequently, the Hilbert space becomes exclusively composed of small fragments, resulting in strong fragmentation.

Notice that before the threshold density $\eta_2$, strong fragmentation can still occur at a system-size dependent value of $\eta^*$, as shown in Fig.\,\ref{fig:figure_0}\textbf{(B)}. This regime is characterized by the coexistence of a single large fragment with numerous small fragments, and the ratio of the largest fragment's size to the total number of states becomes small.
Precisely determining $\eta^*$ requires comparisons over multiple system sizes at the same particle density, a challenging task due to computational constraints in studying large systems. 
However, in the thermodynamic limit, $\eta^*$ is expected to converge to the cutoff beyond which only small fragments exist, eventually approaching $1/2$.

The fragmentation structure plays a crucial role in shaping the dynamical properties of the system, as the dynamics within smaller fragments are tightly constrained and exhibit non-ergodic behavior. 
In contrast, the dynamics within the largest fragment are expected to follow ergodic behavior, unless quantum interference effects lead to deviations from this expectation, inducing non-ergodic dynamics.
In addition to its impact on the dynamical properties, fragmentation also enhances computational efficiency. This phenomenon reduces the computational cost of simulating the dynamics by enabling us to focus solely on the relevant Hamiltonian block corresponding to each initial configuration.
The following sections will explore the dynamics of various types of defects, specifically point-like defects and interfaces, emphasizing the influence of fragmentation on their properties.

\section{Results and Discussion}
In the case of free hard disks \cite{QHD}, we previously investigated the dynamics of defects and observed a strikingly non-ergodic phenomenon in which certain initial defects preserve the underlying crystal structure even for infinitely long times.
In the following section, we will explore in detail the robustness of this phenomenon upon introducing additional soft-core interactions.

\begin{figure*}
    \centering
    \includegraphics[width=\linewidth]{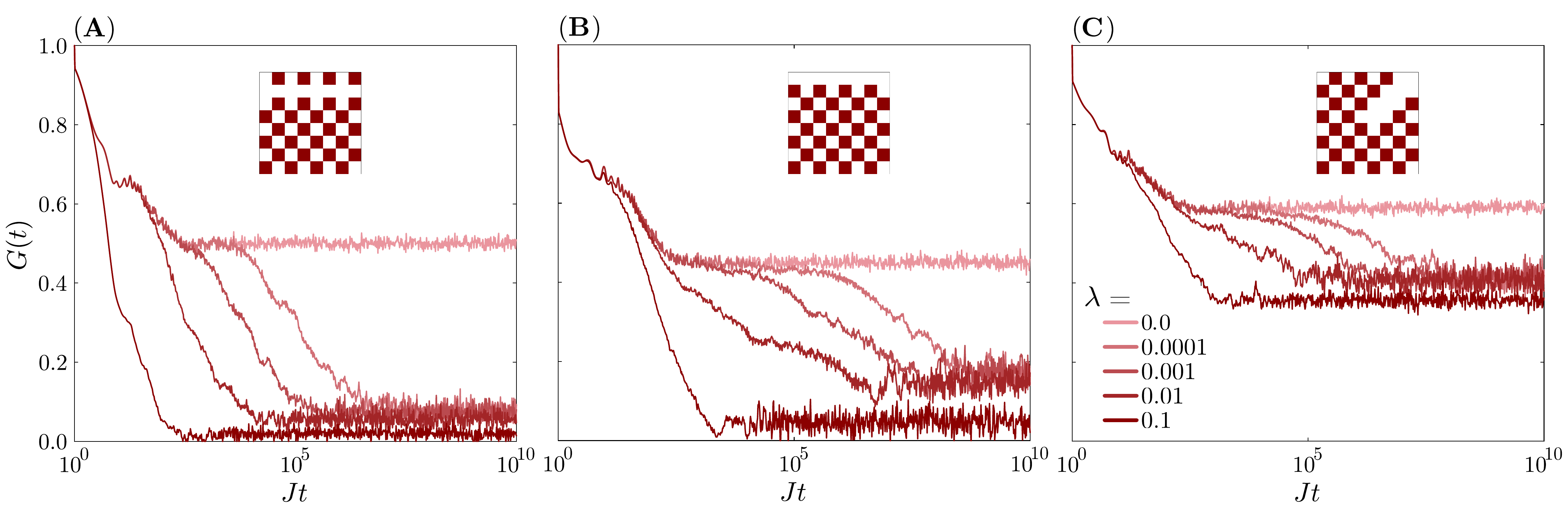}
    \caption{ Memory of the initial condition measured through the autocorrelation function $G(t)$ as a function of the interaction strength $\lambda$ for different initial configurations. The three initial configurations correspond to removing particles in (\textbf{A}) the second row, showing fast relaxation; (\textbf{B}) the first row exhibiting a long plateau; and (\textbf{C}) along half of the diagonal, leading to indefinite memory retention.}
    \label{fig:figure_1}
\end{figure*}

\subsection{Dynamics of point-like defects}
In the following, we will term point-like defects as defects in the crystalline pattern whose size does not depend on the linear dimension of the lattice.
As these are finite-size defects, when the system size increases, a crystal with such a defect has the property that the particle density eventually exceeds the threshold $ \eta > \frac{1}{2} - \frac{\lceil L/2 \rceil}{L^2} $, leading the system to transition into a strongly fragmented regime.
In this regime, only small fragments exist, and within these small fragments, the dynamics remain tightly restricted, effectively preserving the crystalline structure over time.
This has been illustrated by the example shown in Fig.\,\ref{fig:figure_0}\textbf{(C)}, where due to the strong fragmentation point-like defects retain the memory of the initial states.

As a result, initial configurations containing point-like defects exhibit non-ergodic behavior, with memory of the initial state persisting for an infinite time.
Since the introduction of interactions does not alter the fragmentation structure, these defects behave similarly even in the presence of interactions.
Therefore, any point-like defects created in a crystal would preserve its crystal structure indefinitely, despite the presence of interactions.

\subsection{Dynamics of Interfaces}
We will use the notion of interfaces in the following for those type of defects where the number of vacancies in the crystalline pattern depends extensively on the linear dimension of the lattice.
Consequently, such configurations fall within the particle density regime characterized by the coexistence of one large fragment alongside many small fragments. 
Within this regime, interfaces can be classified into two types: those belonging to small fragments and those within the large fragment.
Interfaces in small fragments exhibit non-ergodic behavior and preserve their crystalline structure throughout the dynamics, similar to what is observed for point-like defects.
This behavior persists even in the presence of interactions, as the localized nature of the small fragments constrains the system's evolution.
A simple example of such an interface is formed by removing a middle row of hard disks from the crystal lattice.
This defect configuration becomes completely frozen during the dynamics, preventing any of the hard disks from hopping.

However, the interfaces in the largest fragment are the most intriguing.
Previous work on free hard disks \cite{QHD} has shown that many interface-like defects in the largest fragment exhibit ergodic behavior, losing their crystalline structure over time. 
However, remarkably, certain defects within the large fragment retain their initial crystalline structure indefinitely, even at infinite times.
These results were compared with classical dynamics, where the memory of the initial defect configuration is entirely erased.
This contrast suggests that non-ergodicity in the quantum case arises due to quantum interference effects rather than the fragmentation structure itself.
A key question that remains is how interactions influence this behavior. 
In the remainder of this paper, we focus on precisely this issue, examining how interactions affect the non-ergodic nature of interface-like defects within the large fragment.

We explore the dynamics numerically by varying the interaction strength $\lambda$, system size $L$, and initial configurations with crystal defects. 
Time evolution is performed using the Lanczos algorithm with seven Krylov vectors per time step, with a step size of $0.1$.
Our analysis based on finite-size numerical simulations uncovers three distinct dynamical behaviors associated with different types of initial defects:
\begin{enumerate}
    \item \textbf{Fast relaxation:} Some initial defects lose the memory of the crystal structure in a very short time upon introduction of interactions.  
    \item \textbf{Slow relaxation:} Some defects exhibit remarkable stability over extended periods, but eventually lose the memory of the crystal structure.
    \item \textbf{Retention of initial crystal structure:} Interestingly, certain defects remain intact, even in the presence of interactions and at infinite times.
\end{enumerate}
These findings represent a departure from the dynamical heterogeneity described in \cite{Garrahan2018}, where all initial states—whether characterized by slow or fast relaxation — ultimately thermalize.
In contrast, in our hard-disk model, we observe that certain initial states never undergo thermalization and retain their crystal structure intact.

In this work, we provide representative results for each of these three scenarios, illustrating the dynamics with corresponding initial states. 
To quantify the memory of the crystal pattern over time, we use the auto-correlation function $G(t)$: 
\begin{equation}
    G(t) = \frac{1}{L^2}\sum_{i=1}^{L^2} \left\langle \left(2n_i(t) - 1\right) \left(2n_i(0) - 1\right)\right\rangle - G^*~.
\end{equation}  
\noindent  
Here, $\langle \dots \rangle = \langle \psi | \dots | \psi \rangle$ denotes the expectation value, where $|\psi\rangle$ represents the initial state. A constant $G^* = \left(2\eta - 1\right)^2$ is subtracted to ensure that $G(t) \to 0$ when the initial crystal structure is completely lost, i.e.,  $\langle n_i(t)\rangle \to \eta$. Here, $\eta=M/L^2$ is the particle density given by the ratio between the total number of particles $M$ and system size $L^2$. 

In Fig.\,\ref{fig:figure_1}, we present $G(t)$ as a function of time $t$ for three scenarios with varying interaction strengths $\lambda$. 
The system size considered is $8 \times 8$ with $28$ particles, and the configurations in the inset correspond to the initial states for the respective dynamics.
Notably, an increase in $\lambda$ leads to a decay in $G(t)$ across all cases. In Fig.\,\ref{fig:figure_1}\textbf{(A)}, $G(t)$ rapidly approaches zero, illustrating a scenario where the initial crystal structure quickly dissipates. 
In contrast, Fig.\,\ref{fig:figure_1}\textbf{(B)} shows a much slower decay, with $G(t)$ eventually stabilizing at a small, nonzero value, suggesting partial retention of the crystal structure over time. 
Most intriguingly, Fig.\,\ref{fig:figure_1}\textbf{(C)} reveals that $G(t)$ stabilizes at a significantly higher value, indicating strong preservation of the crystal-like structure despite the presence of interactions.

\begin{figure*}
    \centering
    \includegraphics[width=\linewidth]{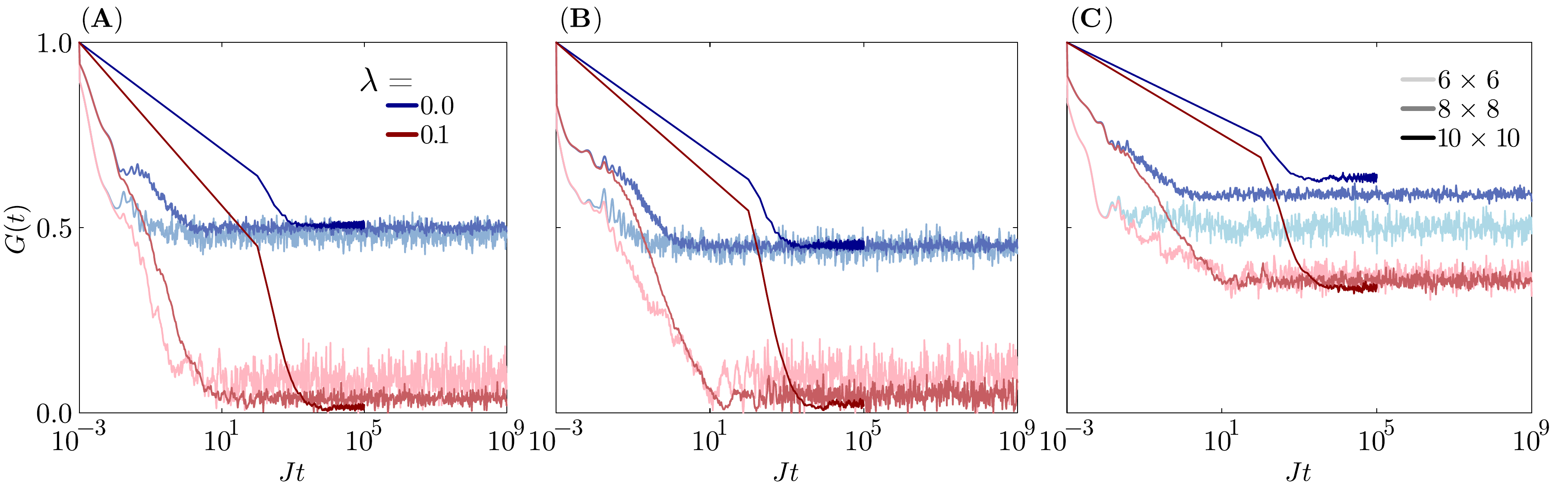}
    \caption{Finite-size dependence of the autocorrelation function $G(t)$ in the interacting and free hard disks corresponding to the initial configurations: (\textbf{A}) particles removed in the second row, (\textbf{B}) particles removed in the first row, and (\textbf{C}) particles removed along half of the diagonal. }
    \label{fig:figure_2}
\end{figure*}
To generalize the analysis, it is essential to investigate the behavior as a function of system size. 
When comparing dynamics across different system sizes, the initial configuration must be designed to ensure that the number of defects scales extensively with the linear dimension of the lattice $L$.
This guarantees that defects form interfaces rather than isolated, point-like structures.
As illustrated in Fig.\,\ref{fig:figure_1}, the initial defect configurations are carefully designed to maintain interface formation across different system sizes. 
Specifically,  the configuration in Fig.\,\ref{fig:figure_1}\,\textbf{(A)} can be systematically scaled by removing particles from the second row. Similarly, the configurations in Fig.\,\ref{fig:figure_1}\,\textbf{(B)} and Fig.\,\ref{fig:figure_1}\,\textbf{(C)}, is scaled by removing particles from the first row and along half of the diagonal, respectively.

In Fig.\,\ref{fig:figure_2}, we present the dynamics with these initial defect configurations for system sizes $L = 6, 8, 10$, with $M = 15, 28, 45$ particles, respectively. 
For free hard disks ($\lambda = 0.0$), $G(t)$ for different system sizes converges to the same value for the initial states shown in Fig.\,\ref{fig:figure_2}\textbf{(A)} and \textbf{(B)}. 
Interestingly, in Fig.\,\ref{fig:figure_2}\textbf{(C)}, $G(t)$ converges to a higher value as the system size increases, suggesting enhanced crystal retention.
For $\lambda = 0.1$, the behavior remains consistent across all system sizes, following the trends discussed earlier for the fixed system size. 
Remarkably, in Fig.\,\ref{fig:figure_1}\textbf{(C)}, $G(t)$ appears to converge to the same value for different system sizes, indicating that certain defects preserve their memory over infinite time, even as the system size increases. 
This observation highlights the robustness of specific initial states, which maintain their crystal structure indefinitely, despite the presence of interactions and finite-size dependence.

\begin{figure*}
    \centering
    \includegraphics[width=\linewidth]{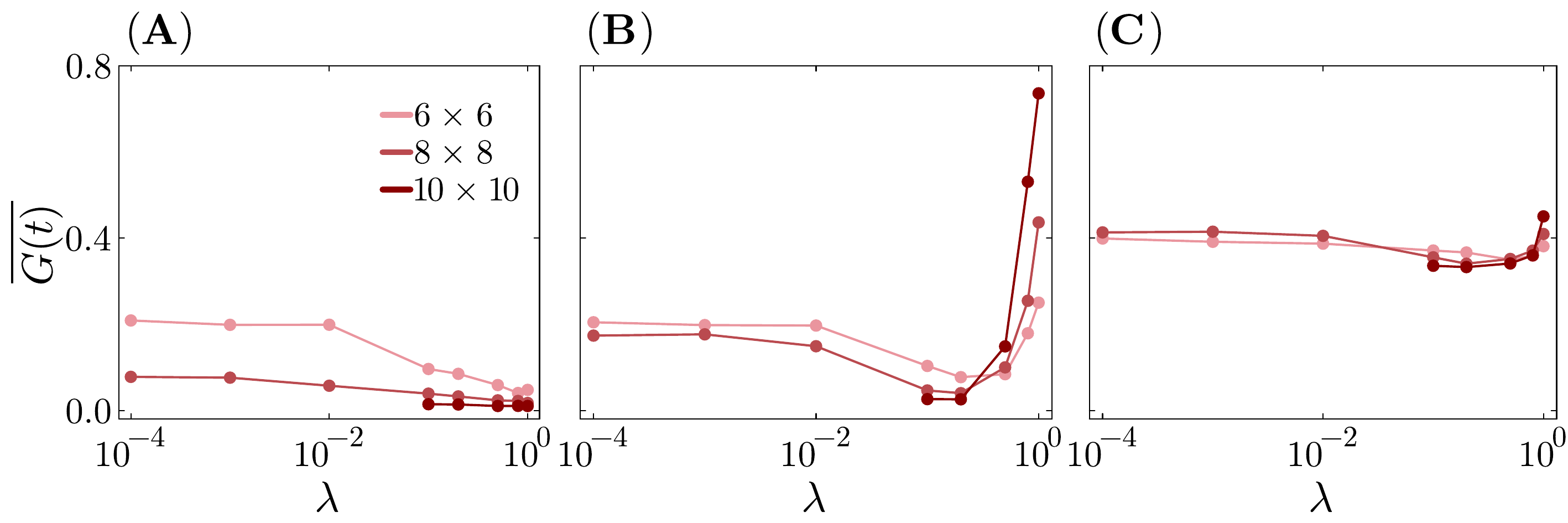}
    \caption{Long-time average of the autocorrelation function $G(t)$ as a function of the interaction strength $\lambda$ corresponding to the initial configurations: (\textbf{A}) particles removed in the second row, (\textbf{B}) particles removed in the first row, and (\textbf{C}) particles removed along half of the diagonal. The time-window considered here is \(Jt=10^4 - 10^5.\)}
    \label{fig:figure_3}
\end{figure*}

To gain a comprehensive understanding of the dynamics across varying interaction strengths and system sizes, we analyze the long-time averaged value of $G(t)$ after convergence, denoted as $\overline{G(t)}$, for several values of $\lambda$ and system sizes. 
The results are presented in Fig.\,\ref{fig:figure_3} for all three scenarios. 
In Fig.\,\ref{fig:figure_3}\textbf{(A)}, $\overline{G(t)}$ consistently decreases with increasing $\lambda$ and system size, indicating rapid structural decay under these conditions. 
Interestingly, Fig.\,\ref{fig:figure_3}\textbf{(B)} reveals a non-monotonic trend where $\overline{G(t)}$ initially decreases with $\lambda$ but increases again for larger $\lambda$, a behavior confirmed across different system sizes. 
In contrast, Fig.\,\ref{fig:figure_3}\textbf{(C)} demonstrates that $\overline{G(t)}$ remains remarkably stable, showing little sensitivity to changes in $\lambda$ or system size.
This stability suggests that certain configurations are inherently robust, maintaining their crystal structure indefinitely despite varying interactions and finite-size dependence.

The increase in $\overline{G(t)}$ for larger $\lambda$ might not always be associated with non-ergodic behavior.
In the limit of large $\lambda$, the next-to-nearest-neighbor interaction becomes the dominant term, with the hopping term as a weak perturbation.
Our initial state is an eigenstate of the interaction term of the Hamiltonian, and the first immediate effect of the hopping term would be to mix the eigenstates of the same energy according to degenerate perturbation theory.
For small systems, there might not be sufficient spreading of the eigenenergy spectrum for such a previously degenerate band, so that different such bands don't yet overlap.
This, therefore might for small systems lead to features of non-ergodic behavior, which, however, would always disappear on increasing system size.
In addition, exploring large values of $\lambda$ is challenging. Since quantum hard disks on the lattice are realized through a Rydberg setup, increasing the next-to-nearest-neighbor interactions inevitably leads to even more distant couplings becoming relevant, complicating the model further. 
For this reason, our analysis in this work should be considered mostly only up to moderate values of $\lambda$.
\begin{figure*}
    \centering
    
\includegraphics[width=\linewidth]{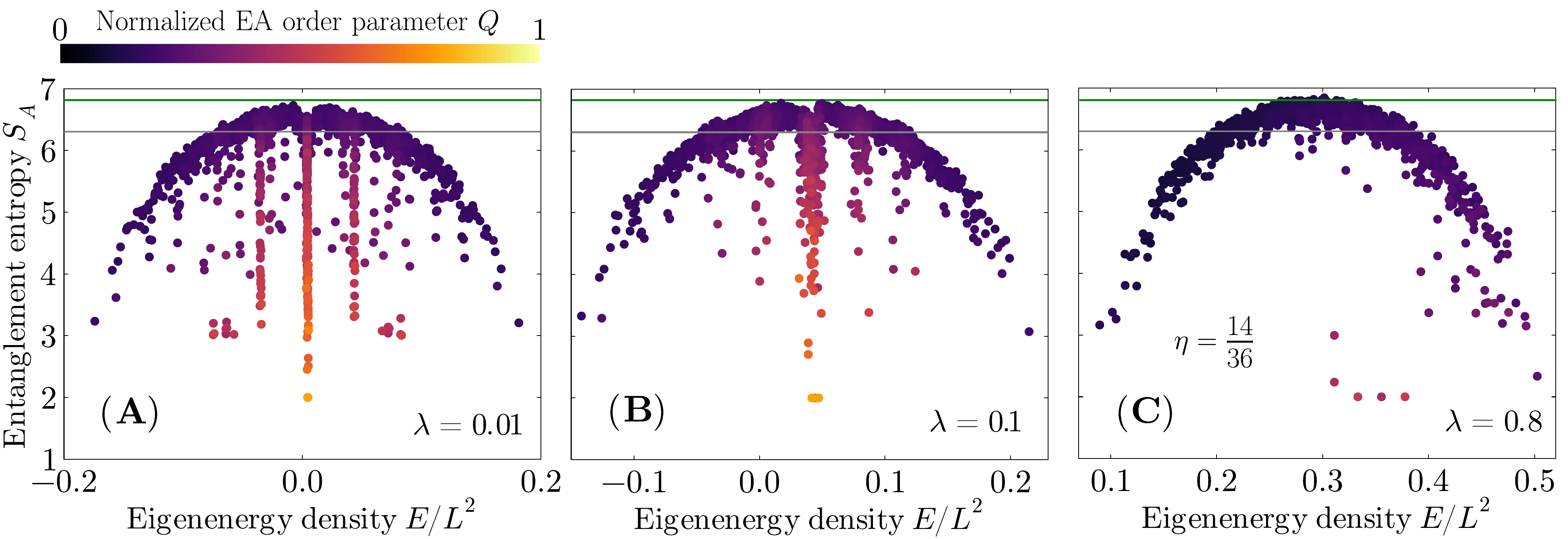}
    \caption{Quantum Many-Body Cages in the presence of interactions. The entanglement entropy of the eigenstates of the largest fragment is shown for different interaction strengths in a $6\times6$ lattice: (\textbf{A}) $\lambda = 0.01$, showing weak interactions; (\textbf{B}) $\lambda = 0.1$, illustrating moderate interactions; and (\textbf{C}) $\lambda = 0.8$, highlighting strong interaction effects. The green line marks the maximal entropy, and the gray line marks the Page entropy. 
Both are computed using the effective subsystem dimension 
$\dim \mathcal{H}_A \approx \sqrt{\dim \mathcal{N}_{\max}}$, 
with the Page value obtained from the standard approximation 
$S_{\mathrm{Page}} \simeq \ln(\dim \mathcal{H}_A) - \tfrac{1}{2}$.}
    \label{fig:figure_4}
\end{figure*}

The remarkable memory retention observed in the third scenario, in the presence of interactions, suggests a connection to atypical eigenstates, similar to those identified for free hard disks \cite{QHD, benami_2025}.
To further explore this phenomenon, we analyze the eigenstate properties using the bipartite entanglement entropy:
\begin{equation}
    S_A = -\Tr \{ \Tr_{\bar{A}} \big(|\Psi \rangle \langle \Psi |\big) \, \textrm{ln}\left[ \Tr_{\bar{A}} \big(|\Psi \rangle \langle \Psi |\big)\right]\}.
\end{equation}
\noindent
Here, $ |\Psi\rangle $ denotes an eigenstate, $ A $ represents the subsystem chosen as the bottom half of the 2D lattice, and $ \bar{A} $ is its complement.
Our Hamiltonian possesses rotational and reflectional symmetries, which give rise to degeneracies. 
As a consequence, the entanglement entropy is not uniquely defined, since any superposition within a degenerate manifold is also an eigenstate with the same energy, leading to different entropy values depending on the chosen superposition.
To eliminate this ambiguity, we work in a symmetry-resolved basis.
Specifically, we apply the full dihedral symmetry group of order four on the $6\times 6$ lattice and diagonalize the Hamiltonian within the $A_1$ irreducible representation.
This procedure enforces the lattice symmetries at the level of the Hilbert space itself and yields uniquely defined eigenstates.

Additionally, to quantify how closely a given eigenstate resembles a specific many-body configuration,  we employ the Edwards-Anderson (EA) order parameter $ Q_{\textrm{EA}} = L^{-4} \sum_{i,j}^{} |\langle \Psi |(2 n_i - 1) (2 n_j - 1 ) |\Psi \rangle |^2. $
We introduce a normalized Edwards--Anderson (EA) order parameter to ensure that it lies between 0 and 1:
\begin{equation}
    Q = \frac{Q_{\mathrm{EA}} - (2\eta - 1)^4}{Q_{\mathrm{EA}}^{\mathrm{max}} - (2\eta - 1)^4}.
\end{equation}
Here, $Q_{\mathrm{EA}}^{\mathrm{max}}$ denotes the maximal value of $Q_{\mathrm{EA}}$ attainable within the $A_1$ irrep subspace.
With this normalization, $Q = 1$ indicates that $|\Psi\rangle$ corresponds to a single basis state of the subspace, whereas $Q = 0$ signifies that $|\Psi\rangle$ is a uniform superposition over all basis states.

In Fig.\,\ref{fig:figure_4}, we present $S_A$ for all eigenstates in the largest fragment in the weak fragmentation region as a function of energy density $E/L^2$ for various interaction strengths $\lambda$, at a fixed particle density $\eta = 14/36$. 
The data points are further distinguished by the normalized EA order parameter $Q$, represented through a color gradient, providing additional insight into the structural characteristics of the eigenstates.
To broaden this investigation beyond the largest fragment, Appendix\,\ref{sec:appendixB} presents Fig.\,\ref{fig:figure_A2}, which displays the same entanglement entropy versus energy density plot, now incorporating contributions from all fragments. This highlights the presence of additional non-ergodic states across different fragments.



In Figs.\,\ref{fig:figure_4}\textbf{(A)}, \textbf{(B)}, and \textbf{(C)}, as expected from the Eigenstate Thermalization Hypothesis, a majority of the eigenstates follow the conventional ergodic paradigm. 
However, intriguingly, certain eigenstates exhibit low entanglement entropy $S_A$ and a high EA order parameter $Q$, even in the middle of the spectrum, deviating from ergodic expectations. 
These states are identified in the quantum hard disk model as quantum many-body cages, a recently developed signature of non-ergodic behavior \cite{benami_2025}.
Remarkably, our results suggest that even with interactions, quantum many-body cages persist.
Although the number of such states decreases as the interaction strength $\lambda$ increases, a finite number of cages remain even for large $\lambda$.
Notice that the prominent towers of eigenstates visible in the spectrum are not exact degeneracies. 
These towers become exactly degenerate only at $\lambda = 0$. 
Once interactions are introduced, the levels split and the degeneracy is lifted.

The presence of these cages directly impacts the dynamics, as initial states with a higher overlap with these eigenstates, such as the one shown in Fig.\,\ref{fig:figure_1}\textbf{(C)}, retain their crystal structure indefinitely.
This phenomenon is broadly accessible within the model and not fine-tuned, as indicated by the consistently high $Q$ values for a substantial number of these states.
Moreover, despite all states belonging to the same fragment, the high $Q$ values observed for cages states indicate quantum interference effects as the underlying cause.
This observation highlights the robustness of non-ergodic behavior in this model, demonstrating that it endures despite the introduction of interactions.

\section{Conclusion and outlook}

In this work, we studied the quantum hard disk model under the influence of interactions, focusing on its dynamical properties and the robustness of quantum many-body cages.
Our results demonstrate that the defining features of the model, such as Hilbert-space fragmentation and non-ergodic dynamics, persist even with the inclusion of interactions.  
The introduction of interactions induces dynamical heterogeneity, with initial defect configurations exhibiting rapid memory loss, long-term decay, or indefinite retention of crystalline structure, depending on the interaction strength and initial configuration.  
Remarkably, quantum many-body cages remain robust in the presence of interactions, enabling certain initial states to retain memory over infinite timescales.
These findings highlight the intricate interplay between constraints and interactions in constrained quantum systems and establish the quantum hard disk model as a versatile platform to explore non-ergodic dynamics and quantum many-body cages.  
Future research could focus on developing a broader framework to understand the robustness of quantum many-body cages due to perturbation.
%
Additionally, it would be valuable to identify the specific properties of initial states that prevent them from thermalizing.  
An intriguing question for future exploration is whether our model's unusual quantum behavior of defects and interfaces extends to other constrained quantum systems beyond hard disks.
Given the fundamental role of interfaces in 2D quantum systems, understanding their stability and dynamics in constrained systems may lead us toward further understanding of non-ergodic behavior.
Furthermore, an important next step would be to move beyond purely numerical approaches and develop an analytical understanding of the stability of interfaces. 
Investigating similar constrained models and identifying practical experimental setups to observe these effects would be crucial for advancing our understanding of non-ergodic quantum matter.

\section*{Acknowledgments} 
We thank Juan Garrahan, Tom Ben-Ami, and Roderich Moessner for insightful discussions.
This project has received funding from the European Research Council (ERC) under the European Union’s Horizon 2020 research and innovation programme (grant agreement No. 853443). 
This work was supported by the German Research Foundation DFG via project 499180199 (FOR 5522) and via project 492547816 (TRR 360).

\section*{Data Availability}
The data to generate all figures in this article is available in Zenodo \cite{Quantum_hard_disks_dataset}.

\nocite{*}
%

\onecolumngrid

\appendix

\section{Classical Dynamics of Free Quantum Hard Disks} \label{sec:appendixA}

In this appendix, we analyze the classical dynamics for the three cases depicted in Fig.\,\ref{fig:figure_2} in the context of free hard disks ($\lambda = 0$).
Our goal is to demonstrate that the classical counterpart of quantum dynamics is ergodic, meaning that the initial crystalline structure is entirely lost over time.
This observation suggests that the non-ergodic behavior found in the quantum case arises due to quantum interference effects rather than the fragmentation structure itself.

\begin{figure*}[h!]
    \centering
    \includegraphics[width=0.9\linewidth]{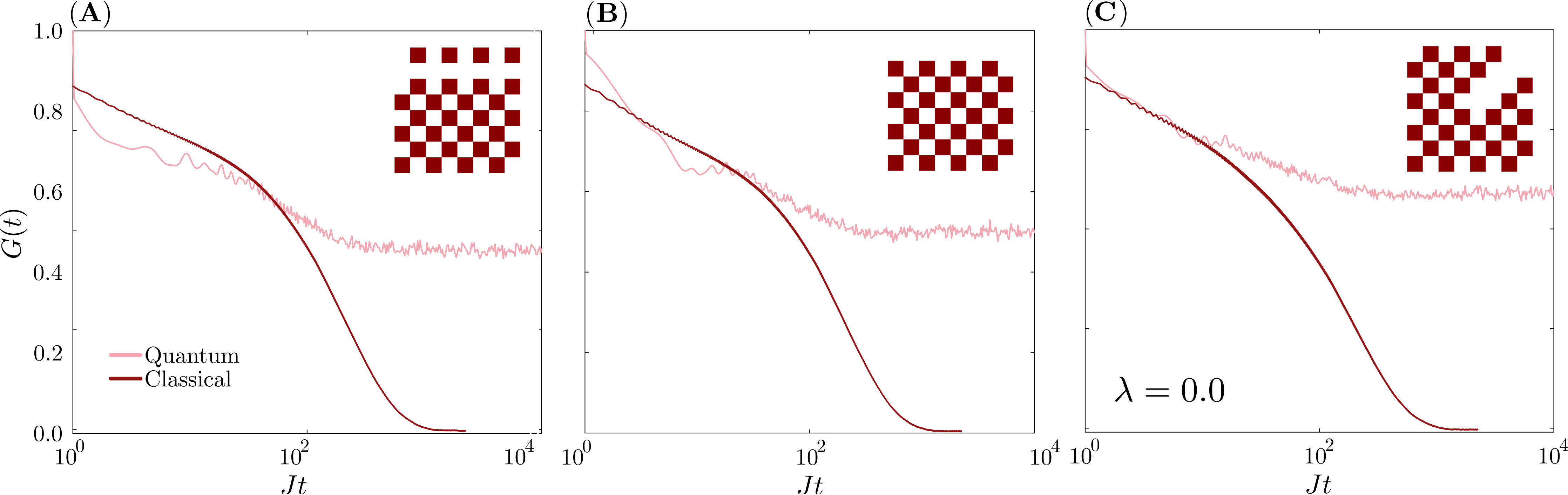}
    \caption{Comparison between quantum and classical dynamics of free hard disks. The autocorrelation function $G(t)$ as a function of the interaction strength $\lambda$ corresponding to the initial configurations: (\textbf{A}) particles removed in the second row, (\textbf{B}) particles removed in the first row, and (\textbf{C}) particles removed along half of the diagonal. }
    \label{fig:figure_A1}
\end{figure*}

The most natural classical analogue of quantum dynamics in this context is a random walk with constraints.
This is motivated by the fact that quantum evolution can be understood as a quantum walk with constraints \cite{QHD}.
As shown in Fig.\,\ref{fig:figure_A1}, the autocorrelation function for classical dynamics approaches zero in all three cases, indicating that the initial crystalline order is completely lost. This confirms the ergodicity of the classical system.

\section{Quantum Many-Body Cages Across the Entire Hilbert Space} \label{sec:appendixB} 
\begin{figure*}[h!]
\includegraphics[width=\linewidth]{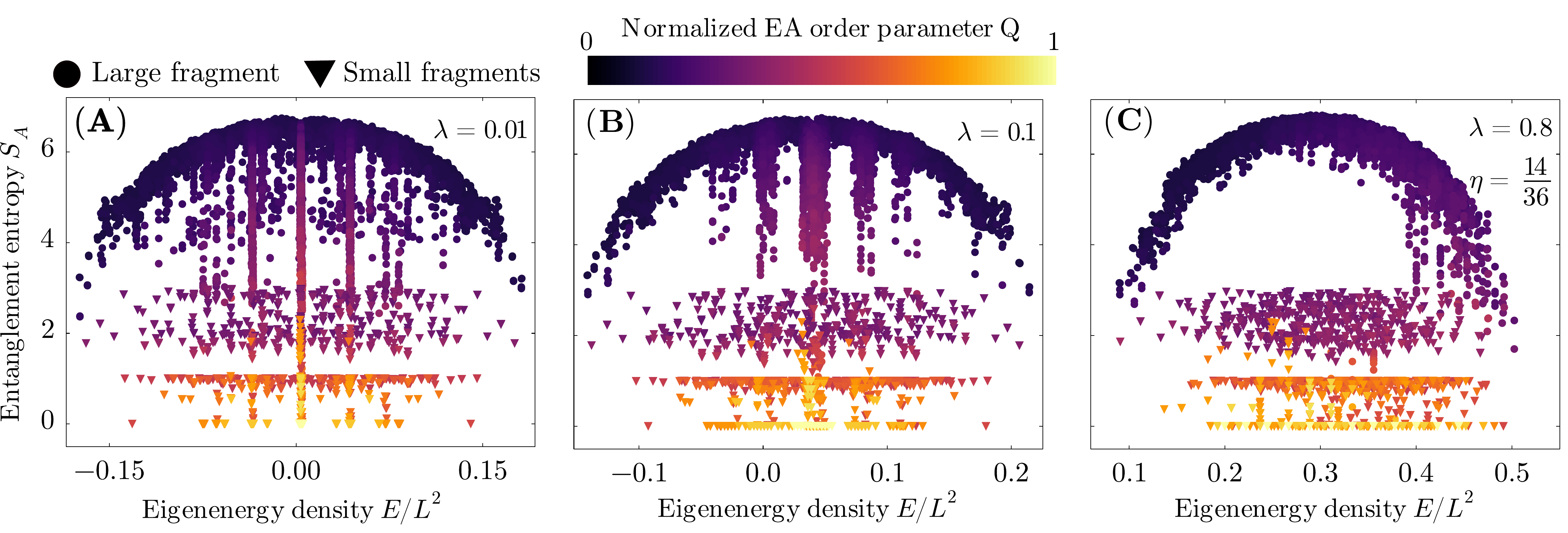}
    \caption{Quantum Many-Body Cages across the entire Hilbert space. The entanglement entropy of the eigenstates of all fragments is shown for different interaction strengths: (\textbf{A}) $\lambda = 0$, representing free hard disks; (\textbf{B}) $\lambda = 0.01$, showing weak interactions; (\textbf{C}) $\lambda = 0.1$, illustrating moderate interactions; and (\textbf{D}) $\lambda = 0.8$, highlighting strong interaction effects.}
    \label{fig:figure_A2}
\end{figure*}

Fig.\,\ref{fig:figure_4} illustrates the entanglement entropy as a function of eigenenergy density, with data points colored by the normalized EA order parameter. 
This visualization specifically focuses on the largest fragment within the weak fragmentation regime. 
To extend this analysis, Fig.\,\ref{fig:figure_A2} presents the same plot of entanglement entropy versus eigenenergy density for all fragments.
This reveals that numerous additional cages emerge from the smaller fragments, complementing those found in the largest fragment. 
Notably, the cages observed in the largest fragment primarily stem from quantum interference effects, whereas those in the smaller fragments emerge due to the fragmentation structure itself. 

\end{document}